\documentclass[twocolumn,showpacs,preprintnumbers,amsmath,amssymb,prl,superscriptaddress]{revtex4-1}
\usepackage{bbm}
\usepackage{mathrsfs}
\usepackage{graphicx}
\usepackage{dcolumn}
\usepackage{bm}
\usepackage{amsmath}
\usepackage{amsfonts}
\usepackage{color}
\usepackage{floatrow}

\begin{document}

\title{Magnetic Dirac Semimetals in Three Dimensions}
\author{Jing Wang}
\affiliation{State Key Laboratory of Surface Physics and Department of Physics, Fudan University, Shanghai 200433, China}
\affiliation{Collaborative Innovation Center of Advanced Microstructures, Nanjing 210093, China}

\begin{abstract}
We present a new type of three-dimensional essential Dirac semimetal with magnetic ordering.
The Dirac points are protected by the magnetic space groups and cannot be gapped without lowering such symmetries, where the combined antiunitary symmetry of half-translation operator and time-reversal plays an essential role. We introduce two explicit tight-binding models for space groups~$16$ and $102$, which possesses Dirac point at time-reversal-invariant momenta of surface Brillouin zone. In contrast to the time-reversal-invariant essential Dirac semimetal, the magnetic space groups here can be either symmorphic or non-symmorphic, and the magnetic DSM is symmetry tuned to the boundary between weak topologically distinct insulating phases. Interestingly, the symmetry-breaking perturbations could lead to an ideal Weyl semimetal phase with only two minimal Weyl points pinned exactly at the Fermi energy for filling $\nu\in4\mathbb{Z}+2$. By reducing the dimensionality we are able to access the Dirac and Weyl semimetal phases in two dimensions.
\end{abstract}

\date{\today}

\maketitle

The experimental discovery of the time-reversal invariant topological insulators (TIs)~\cite{hasan2010,qi2011} has inspired intense research interest in the symmetry-protected topological phases of matter. Recently, great attention has been given to the topological semimetals in various spatial dimensions, which have nontrivial surface states (SSs) and may be characterized by Fermi surface topological invariants~\cite{volovik2003,horava2005,zhao2013,lian2016}.
The three-dimensional (3D) semimetals can be classified in terms of degeneracies of the crossing points in the electronic structure, which includes Dirac semimetal (DSM), Weyl semimetal (WSM), Double DSM and Spin-$1$ WSM~\cite{young2012,wangzj2012,wangzj2013,young2014,yang2014,liu2014a,liu2014b,borisenko2014,xiong2015,murakami2007,wan2011,xu2011,burkov2011,liu2014,huang2015,weng2015,xu2015,lv2015,bernevig2015,sun2015,ruan2016,deng2016,wieder2016,bradlyn2016}.
Among them, DSM is particularly interesting because it is the parent state of various exotic quantum states such as TI and WSM. The DSM hosts massless Dirac fermions with linear energy dispersions as the low-energy excitation, where the conduction and valence bands contact only at the Dirac points (DPs) in the Brillouin zone (BZ). The DSM materials have both the time-reversal ($\Theta$) and inversion ($\mathcal{I}$) symmetries~\cite{young2012,wangzj2012,wangzj2013,young2014,yang2014,liu2014a,liu2014b,borisenko2014,xiong2015}, which fall into two distinct classes. The first class is the \emph{topological} DSM such as Cd$_3$As$_2$ which is induced by band inversion and locally permitted by crystalline symmetries, while the second class is the \emph{essential} DSM such as BiO$_2$ where the nodal features are filling-enforced by specific space group (SG) symmetries~\cite{parameswaran2013,watanabe2015,po2016,watanabe2016,wieder2016b}.

The $\Theta$ breaking in general splits a DP into Weyl points and a WSM is obtained. The concept of magnetic DSM in 2D has been introduced recently~\cite{wang2016c,young2016}. The goal in this paper is to explain how a magnetic DSM with broken $\Theta$ symmetry can nevertheless exist in 3D. Quite different from the antiferromagnetic (AFM) DSM considered in Ref.~\cite{tang2016}, where
DPs are created through band inversion and belongs to the first class. Here the magnetic DSM is \emph{essential}, where the bulk DP is protected by (either symmorphic or non-symmorphic) magnetic SG and cannot be gapped without lowering such symmetries. An essential role is played by the combined symmetry of $\Theta$ and translation operator, namely $\mathcal{S}=\Theta T_{\mathbf{d}}$, where $\mathbf{d}$ is half of a primitive-lattice vector. We introduce two explicit tight-binding models for SGs~$16$ and $102$, which possesses DP at time-reversal-invariant momenta (TRIM) of surface BZ with $\mathcal{S}^2=-1$. Like $\Theta$-invariant essential DSM, the magnetic DSM is symmetry tuned to the boundary between topologically distinct insulating phases. We conclude with a brief discussion of the measurable consequences and possible material venues for these phases.

In a 3D $\Theta$-invariant system, DSM emerges at the transition between a normal insulator (NI) and a TI or weak TI/topological crystalline insulator (TCI)~\cite{murakami2007,yang2014}. The DP is further stabilized by crystallographic symmetries, where Kramers degeneracy from $\Theta$ plays a key role. In a 3D $\Theta$-broken magnetic system, a new $Z_2$ invariant can be defined in a 2D BZ when $\mathcal{S}$-symmetry is present, separating a NI from AFM TI~\cite{mong2010,fang2013}. Therefore, a magnetic DSM is expected to arise at the phase boundary.
To have an intuitive picture, we first consider the sufficient condition for the existence of a fourfold degeneracy at certain $\mathbf{k}$ in the BZ. For high symmetry $\mathbf{k}$ points which are left invariant under spatial operations $\mathcal{G}_1$ and $\mathcal{G}_2$, if for example, $\mathcal{G}_1,\mathcal{G}_2$ satisfy $\{\mathcal{G}_1,\mathcal{G}_2\}=0$ and $\mathcal{G}_1^2=\pm1$, $\mathcal{G}^2_2=\pm1$, there exists a twofold degeneracy. The $\Theta$ symmetry will set constraints at TRIM. If $\{\Theta,\mathcal{G}_1\}=\{\Theta,\mathcal{G}_2\}=0$, there exits a fourfold degenracy.
This is evident by considering an eigenstate of $\mathcal{G}_1$ satisfying $\mathcal{G}_1\psi=\lambda\psi$, then $\mathcal{G}_2\psi$, $\Theta\psi$ and $\Theta\mathcal{G}_2\psi$ are also eigenstates of $\mathcal{G}_1$ with eigenvalues $-\lambda$, $-\lambda^*$ and $\lambda^*$. All of these eigenstates are orthogonal to each other due to different eigenvalues and antiunitary $\Theta$. In fact, this is the case for DSM predicted in BiZnSiO$_4$ where $\mathcal{G}_1=\{M_{\hat{z}}|0\frac{1}{2}\frac{1}{2}\}$ and $\mathcal{G}_2=\{\mathcal{I}|\frac{1}{2}\frac{1}{2}0\}$~\cite{young2014}.

Now in a $\Theta$-broken but $\mathcal{S}$-invariant system, $\mathcal{S}$ is antiunitary like $\Theta$, and the Hamiltonian is $\mathcal{S}$-invariant as $\mathcal{S}_{\mathbf{k}}\mathcal{H}(\mathbf{k})\mathcal{S}_{\mathbf{k}}^{-1}=\mathcal{H}(-\mathbf{k})$. However, there is a key difference: while $\Theta^2=-1$ for the spin-$1/2$ system, $\mathcal{S}^2\equiv\mathcal{S}_{-\mathbf{k}}\mathcal{S}_{\mathbf{k}}=\Theta^2T_{2\mathbf{d}}=-T_{2\mathbf{d}}$. Therefore, $\mathcal{S}^2=-1$ only at the TRIM satisfying $\mathbf{k}\cdot\mathbf{d}=n\pi$ where the Kramers' degeneracy is preserved, but $\mathcal{S}^2=+1$ at $\mathbf{k}\cdot\mathbf{d}=(n+\frac{1}{2})\pi$ where the bands are generically \emph{nondegenerate}. Therefore, a fourfold degeneracy is guaranteed to exist at the $\mathbf{k}$ point with $\mathbf{k}\cdot\mathbf{d}=n\pi$, if either
\begin{equation}\label{AFM_1}
\{\mathcal{G}_1,\mathcal{G}_2\}=\{\mathcal{S},\mathcal{G}_1\}=\{\mathcal{S},\mathcal{G}_2\}=0,
\end{equation}
with $\mathcal{G}_1^2=\mathcal{G}_2^2=\pm1$; or
\begin{equation}\label{AFM_2}
\{\mathcal{S},\mathcal{G}_1\}=[\mathcal{S},\mathcal{G}_2]=0,\ \ \mathcal{G}_1\mathcal{G}_2\mathcal{G}_1=\mathcal{G}_2,
\end{equation}
with $\mathcal{G}_1^4=-1$ the fourfold rotation.

\begin{figure}[t]
\begin{center}
\includegraphics[width=3.1in]{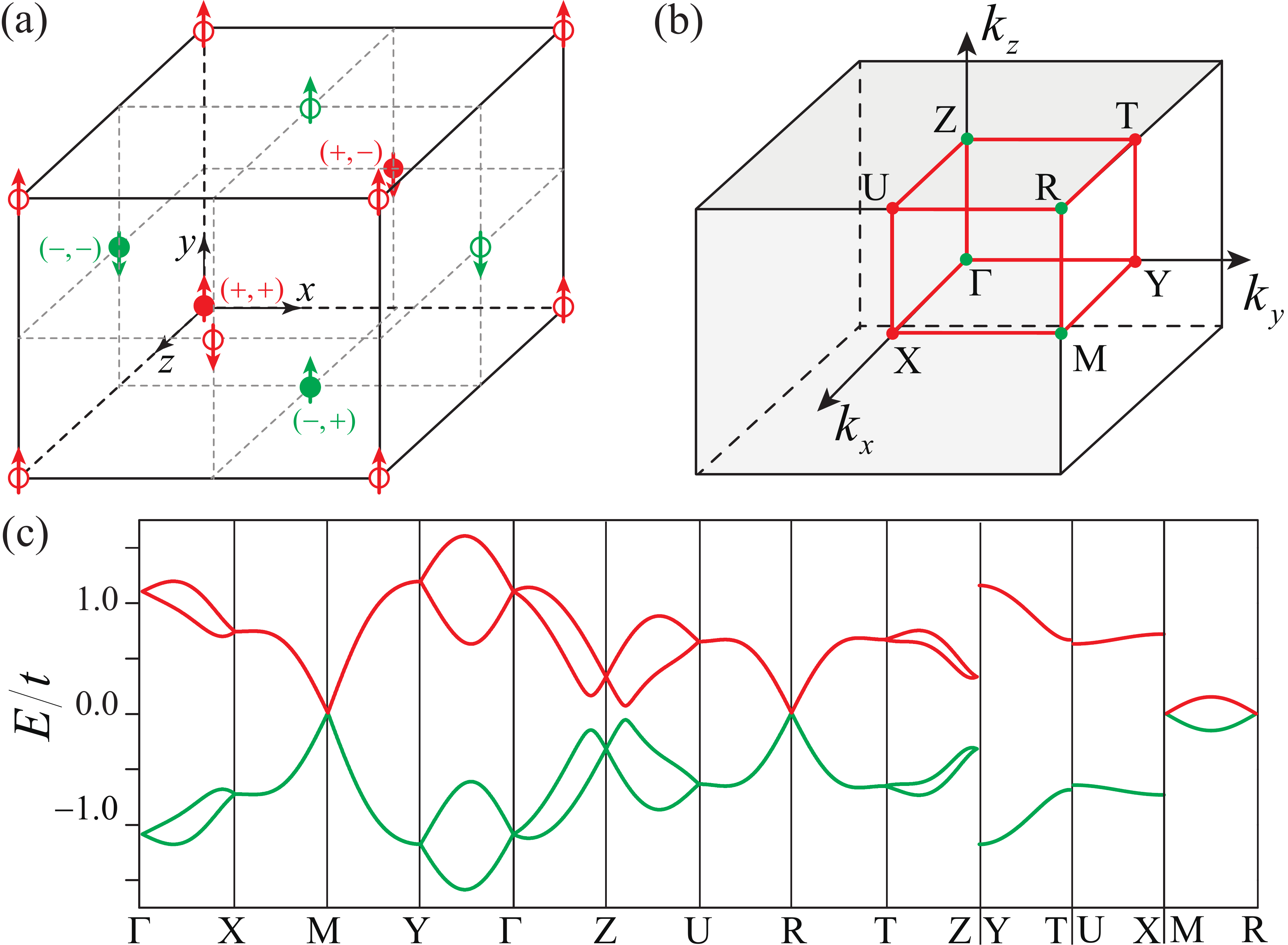}
\end{center}
\caption{(color online). (a) The lattice for the orthorhombic primitive structure of SG~16. The four sublattices are denoted as solid circles and labeled by $(\tau_z,\sigma_z)=(\pm1,\pm1)$. The magnetic moments are along $\pm\hat{y}$ direction. (b) BZ. The green solid circles are TRIM with $\mathbf{k}\cdot\mathbf{d}=n\pi$. (c) Energy band of the AFM system in (a), which is described by the model in Eq.~(\ref{AFM_model1}), with $t=1.0$, $\lambda_1=0.25$, $\lambda_2=0.6$, $\lambda_3=0.4$, $\lambda_4=0.2$, $\lambda_5=0.4$, and $\lambda_6=0.15$. Two DPs at $M$ and $R$ are protected by the magnetic SG in Eq.~(\ref{AFM_1}). Here only the upper four bands of an eight-band model is shown. The band along $M$-$R$ is dispersionless when $\lambda_6=0$.}
\label{fig1}
\end{figure}

\emph{Model A.}
We now study an explicit tight-binding model for a 3D magnetic DSM to illustrate the symmetry-protected DPs listed in Eq.~(\ref{AFM_1}). The lattice has an orthorhombic primitive structure of SG~16 (Shubnikov group $P2'2'2$) as shown in Fig.~\ref{fig1}(a). The lattice vectors are $\vec{\mathbf{a}}_1=(100)$, $\vec{\mathbf{a}}_2=(010)$, $\vec{\mathbf{a}}_3=(001)$. The system has a layered structure with four sublattices in one unit cell, indexed by $(\tau_z,\sigma_z)=(\pm1,\pm1)$ associated with the basis vectors $\mathbf{t}_0=-\tau_z\sigma_z(\frac{1}{4}00)-\sigma_z(0\frac{1}{4}0)-\tau_z(00\frac{1}{4})$. The symmetry generators and their representations in the sublattice space are $\{C_{2\hat{x}}|000\}=\tau_x\sigma_x$ and $\{C_{2\hat{y}}|000\}=\tau_x\sigma_z$. Each lattice site contains an $s$ orbital, which in general leads to an eight-band model. We then introduce the AFM ordering along $\pm\hat{y}$ direction, thus the system respects $\mathcal{S}$ symmetry associated with $\mathbf{d}=(\frac{1}{2}\frac{1}{2}0)$, which is represented by $\mathcal{S}=e^{i\mathbf{k}\cdot\mathbf{d}}i\sigma_y\mathcal{K}$, and $\mathcal{K}$ is complex conjugation. We further assume AFM interaction is much stronger than hopping and spin-orbit coupling (SOC) terms, therefore the system is decoupled into two four-band models as time-reversal partners, each with one spin per sublattice. The upper subsystem consists of $|\uparrow_{++}\rangle$, $|\uparrow_{-+}\rangle$, $|\downarrow_{+-}\rangle$ and $|\downarrow_{--}\rangle$, with $\uparrow$ and $\downarrow$ denoting the spin up and down states, respectively. Here, we consider a generic but simplified model which respects all the symmetries and are sufficient to characterize all the essential degeneracies of the band structure. The Hamiltonian is
\begin{eqnarray}\label{AFM_model1}
\mathcal{H}_{\text{a}} &=& t\tau_x\cos\frac{k_x}{2}\cos\frac{k_z}{2}+\lambda_1\tau_y\cos\frac{k_x}{2}\sin\frac{k_z}{2}
\nonumber
\\
&&+\lambda_2\tau_z\sigma_x\cos\frac{k_x}{2}\sin\frac{k_y}{2}+\lambda_2\sigma_x\sin\frac{k_x}{2}\cos\frac{k_y}{2}
\nonumber
\\
&&+\lambda_3\tau_y\sigma_y\cos\frac{k_y}{2}\cos\frac{k_z}{2}+\lambda_4\tau_x\sigma_y\cos\frac{k_y}{2}\sin\frac{k_z}{2}
\nonumber
\\
&&+\lambda_5\sigma_z\cos k_x\sin k_y+\lambda_5\tau_z\sigma_z\sin k_x\cos k_y
\nonumber
\\
&&+\lambda_6\tau_z\sigma_y\sin\frac{k_x}{2}\sin\frac{k_y}{2}\sin k_z.
\end{eqnarray}
Here $t$ describes the nearest neighbor hopping, where we fix the gauge within the sublattice such that $\mathcal{H}_a(\mathbf{k}+\mathbf{G})=e^{-i\mathbf{G}\cdot\mathbf{t}_0(\tau_z,\sigma_z)}\mathcal{H}_a(\mathbf{k})e^{i\mathbf{G}\cdot\mathbf{t}_0(\tau_z,\sigma_z)}$.
$\lambda_{i}$ is SOC which involves spin-dependent nearest ($i=1,2,3,4$), third nearest ($i=5$), and fourth nearest ($i=6$) neighbor hopping.

Fig.~\ref{fig1}(c) shows energy band associated with $\mathcal{H}_a$, which features two inequivalent DPs at $M$ and $R$ with linear dispersion along all directions. Moreover, we find that there are two additional Weyl points along the line $Z$-$U$ when $2(\lambda_1/\lambda_2)(\lambda_5/\lambda_4)=(\lambda_1/\lambda_2)^2+1$. These two DPs at $M$ and $R$ are protected by $\mathcal{S}$, $\{C_{2\hat{x}}|000\}$ and $\{C_{2\hat{y}}|000\}$, which can be seen by examining the effective model near these points. However, they need not be at the same energy. The representations of symmetry operations at $M$ are, $\{C_{2\hat{x}}|000\}=i\tau_x\sigma_y$, $\{C_{2\hat{y}}|000\}=i\tau_y$ and $\mathcal{S}=i\tau_z\sigma_y\mathcal{K}$. Therefore, the generic $\mathbf{k}\cdot\mathbf{p}$ Hamiltonian at $M$ is
\begin{eqnarray}
\mathcal{H}_{\text{a}}^M &=& (u_1\tau_x-u_2\tau_z\sigma_x+u_3\tau_z\sigma_z)k_x
\nonumber
\\
&&+(v_1\tau_y\sigma_y+v_2\sigma_z+v_3\sigma_x)k_y+w_1\tau_z\sigma_yk_z,
\end{eqnarray}
which leads to the dispersion
\begin{equation}
E_{a,M\pm}^2(\mathbf{k}) =|\mathbf{u}|^2k_x^2+|\mathbf{v}|^2k_y^2+|\mathbf{w}|^2k_z^2\pm2k_xk_y|\mathbf{u}\times\mathbf{v}|.
\end{equation}
where $\mathbf{u}\equiv(u_1,u_2,u_3)$, $\mathbf{v}\equiv(v_1,v_2,v_3)$ and $\mathbf{w}\equiv(w_1,0,0)$. When $\mathbf{u}\perp\mathbf{v}$, i.e., $|\mathbf{u}\times\mathbf{v}|=|\mathbf{u}|\times|\mathbf{v}|$, one of the branches vanishes on the line $|\mathbf{u}|k_x=|\mathbf{v}|k_y$, $k_z=0$. From the tight-binding model, we have $\mathbf{u}=(-\frac{t}{2},\frac{\lambda_2}{2},\frac{\lambda_5}{2})$, $\mathbf{v}=(-\frac{\lambda_3}{2},\frac{\lambda_5}{2},-\frac{\lambda_2}{2})$, so as long as $t\lambda_2\neq0$, there will be a symmetry-protected DP at $M$. Likewise, at $R$, the representations of symmetry operations are, $\{C_{2\hat{x}}|000\}=i\tau_y\sigma_y$, $\{C_{2\hat{y}}|000\}=i\tau_x$ and $\mathcal{S}=i\sigma_y\mathcal{K}$. The $\mathbf{k}\cdot\mathbf{p}$ Hamiltonian at $R$ is
\begin{eqnarray}
\mathcal{H}_{\text{a}}^R &=& (u'_1\tau_y-u'_2\tau_z\sigma_x+u'_3\tau_z\sigma_z)k_x
\nonumber
\\
&&+(-v'_1\tau_x\sigma_y+v'_2\sigma_z+v'_3\sigma_x)k_y+w'_1\tau_z\sigma_yk_z.
\end{eqnarray}
This leads to the dispersion $E_{a,R\pm}^2(\mathbf{k}) =|\mathbf{u}'|^2k_x^2+|\mathbf{v}'|^2k_y^2+|\mathbf{w}'|^2k_z^2\pm2k_xk_y|\mathbf{u}'\times\mathbf{v}'|$, where from the model~(\ref{AFM_model1}), we have $\mathbf{u'}\equiv(u_1',u_2',u_3')=(-\frac{\lambda_1}{2},\frac{\lambda_2}{2},\frac{\lambda_5}{2})$, $\mathbf{v}\equiv(v_1',v_2',v_3')=(\frac{\lambda_4}{2},\frac{\lambda_5}{2},-\frac{\lambda_2}{2})$ and $\mathbf{w}\equiv(w_1',0,0)$, so there are no symmetry respecting terms at $R$ that could lift the degeneracy.

\begin{table}[t]
\caption{Perturbations to the DP of \emph{Model A} in SG~16, classified by their symmetry under $D_2$ point group~\cite{bradley1972}. The resulting insulating and semimetallic phases (with \# of nodes) as a function of the increased perturbation strength are indicated.}
\begin{center}\label{table1}
\renewcommand{\arraystretch}{1.2}
\begin{tabular*}{3.3in}
{@{\extracolsep{\fill}}cccc}
\hline
\hline
Reps & Perturbations & Phases & \# of Nodes
\\
\hline
$A_1$ & $1$ & DSM & 2
\\
$B_1$ & $\tau_z$ & WSM$\rightarrow$NI & $4\rightarrow0$
\\
      & $\tau_y\sigma_x,\tau_x\sigma_x$ & WSM & $4$
\\
      & $\tau_y\sigma_z,\tau_x\sigma_z$ & WSM & $4$
\\
$B_2$ & $\tau_x,\tau_y$ & WSM$\rightarrow$AFM TCI & $4\rightarrow2\rightarrow0$
\\
$B_3$ & $\tau_y\sigma_y,\tau_x\sigma_y$ & WSM$\rightarrow$AFM TCI & $4\rightarrow2\rightarrow0$
\\
\hline
\hline
\end{tabular*}
\end{center}
\end{table}

\begin{figure}[b]
\begin{center}
\includegraphics[width=3.1in]{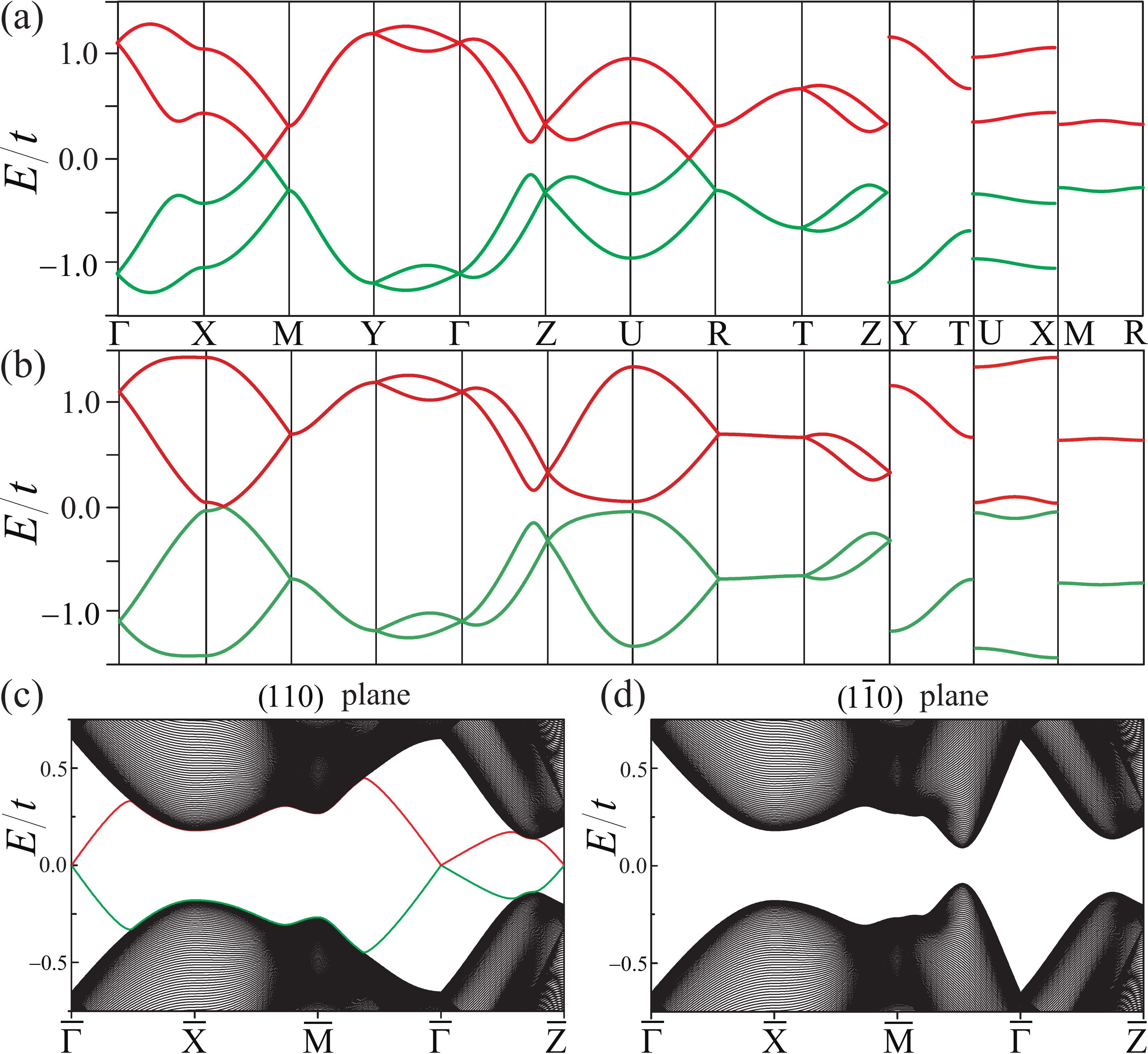}
\end{center}
\caption{(color online). Symmetry breaking phases of \emph{Model A}. (a) Bulk band when $\mathcal{H}_2^a$ is added with $\gamma_2=0.3$. (b) Minimal and ideal WSM with $\gamma_2=0.7$. (c) \& (d) Energy dispersion for a slab configuration with $\gamma_2=0.9$ on (110) \& ($1\bar{1}0$) plane. The (110) plane has gapless Dirac SSs with DPs located at surface TRIM $\bar{\Gamma}$ and $\bar{Z}$.}
\label{fig2}
\end{figure}

We further consider the symmetry-breaking perturbations which may lead to a wealth
of topological phases. The general $\mathcal{S}$-invariant perturbations are listed in Table~\ref{table1}.
Adding these mass terms results in either insulating or WSM phases. Take $\tau_z$ for example, it breaks both $\{C_{2\hat{x}}|000\}$ and $\{C_{2\hat{y}}|000\}$ which allows a term $\mathcal{H}^a_1=\gamma_1\tau_z$. It corresponds to a staggered on-site potential, which results in WSM when $\gamma_1$ is small. Each DP splits into a pair of Weyl points located along the line $M$-$R$, and strong $\gamma_1$ will further push the Weyl points annihilating pairwise, resulting in a NI without gapless SSs, regardless of surface termination~\cite{supplementary}. More interestingly, we consider a lattice distortion which breaks $\{C_{2\hat{x}}|000\}$ but preserves $\{C_{2\hat{y}}|000\}$. Such reduced symmetry adds a term $\mathcal{H}^a_2=\gamma_2\sin(k_x/2)[\sin(k_z/2)\tau_x+\cos(k_z/2)\tau_y]$. When $\gamma_2$ is small, the DP at $M$ ($R$) splits into a pair of Weyl points located along the line $M$-$X$ ($R$-$U$) as shown in Fig.~\ref{fig2}(a). With increased $\gamma_2$, the Weyl points along $U$-$R$ annihilate pairwisely, and the system becomes a \emph{minimal} and \emph{ideal} WSM with only two Weyl points pinned to the Fermi energy for filling $\nu\in4\mathbb{Z}+2$. These fillings are allowed due to absence of multiple nonsymmorphic symmetries. With further increased $\gamma_2$, these two Weyl points meet and annihilate at $X$, resulting in a bulk insulating phase. Such insulating phase is a AFM TCI with trivial $Z_2$ index $\nu_0=0$, which has gapless SSs on certain $\mathcal{S}$-invariant surfaces such as $(110)$ plane. However, it is topological in a weaker sense than strong TI because these SSs are not generally immune to disorder.

\begin{figure}[t]
\begin{center}
\includegraphics[width=3.1in]{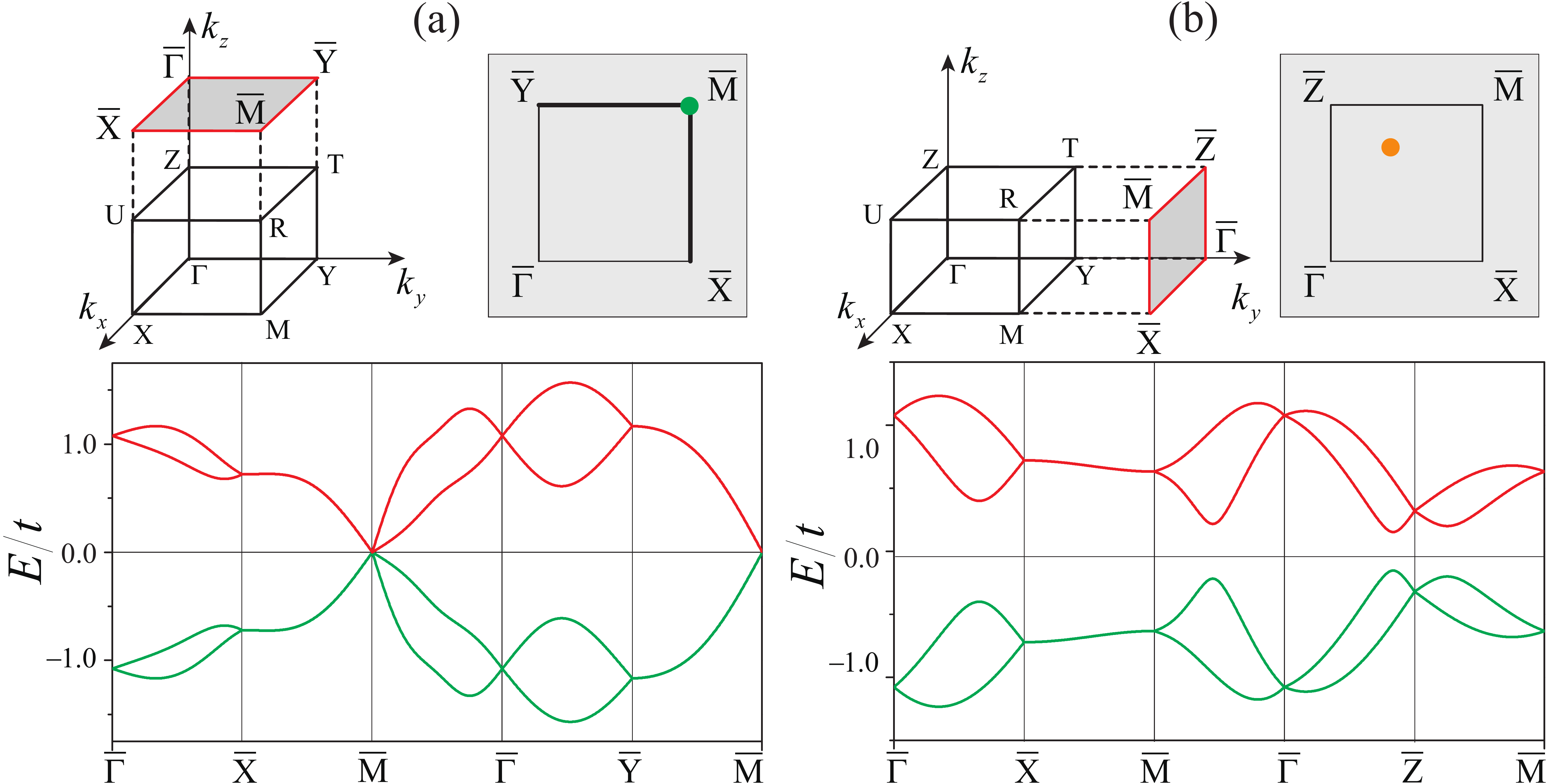}
\end{center}
\caption{(color online). Energy bands for \emph{Model A} in the thin film limit, with 2D BZ shown above.  (a) The bilayer structure along [001] direction. The single DP at $\bar{M}$ is marked as a green dot.
(b) The bilayer structure along [010] direction preserves $\{C_{2\hat{x}}|00\}$ and $\{C_{2\hat{y}}|00\}$ but breaks $\mathcal{S}$, the DPs split into Weyl points (marked as yellow dots).}
\label{fig3}
\end{figure}

We close this section by studying how the 3D DP evolves by reducing dimensionality. The energy band for a [001]-oriented thin film is shown in Fig.~\ref{fig3}(a). The 2D lattice respects $\mathcal{S}$, $\{C_{2\hat{x}}|00\}$ and $\{C_{2\hat{y}}|00\}$, which protects a single 2D DP at $\bar{M}$ following the algebra in Eq.~(\ref{AFM_1}). In fact, the 3D lattice is viewed as a magnetic layer group along [001] direction~\cite{bradley1972}. Therefore, the magnetic crystalline symmetry could protect the essential DP in both 3D and 2D~\cite{young2016}. While for a [010]-oriented thin film in Fig.~\ref{fig3}(b), the lattice preserves $\{C_{2\hat{x}}|00\}$ and $\{C_{2\hat{y}}|00\}$ but breaks $\mathcal{S}$, the DP is no longer protected but splits into 2D Weyl points~\cite{supplementary}. This is in sharp contrast to topological DSM in the first class, where gapped quantum spin Hall state can be obtained by dimension reduction~\cite{wangzj2013}.

\begin{figure}[b]
\begin{center}
\includegraphics[width=3.2in]{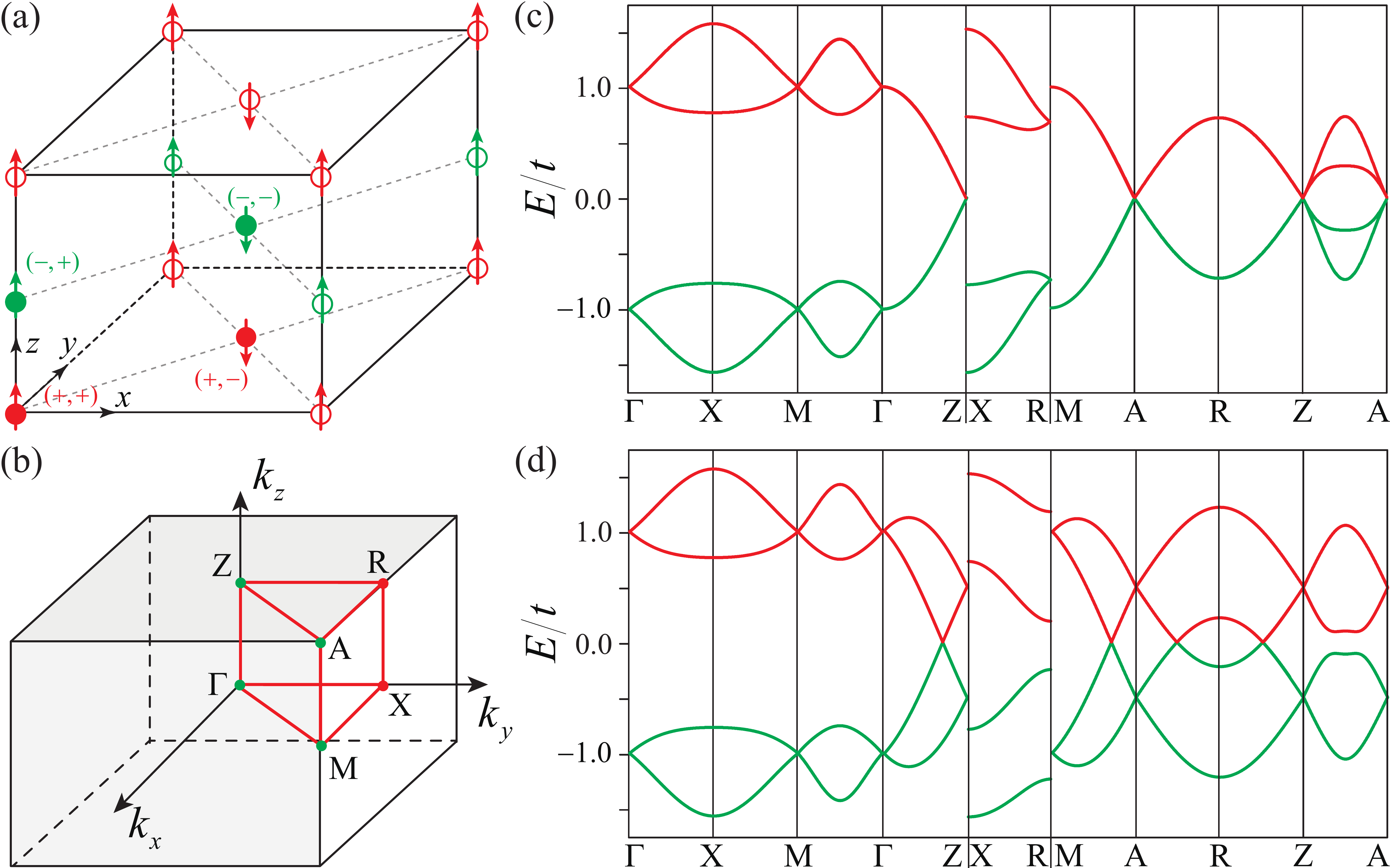}
\end{center}
\caption{(color online). (a) Model lattice for the common tetragonal primitive structure of SG~102. The four sublattices in a unit cell are denoted as solid circles and labeled by $(\tau_z,\sigma_z)=(\pm1,\pm1)$. The magnetic moments are along $\pm\hat{z}$ direction. (b) BZ. The green dots are TRIM with $\mathbf{k}\cdot\mathbf{d}=n\pi$. (c) Energy band for SG~102, which is described by the tight-binding model of Eq.~(\ref{AFM_model2}), with $t=1.0$, $t_1=0.2$, $\lambda_1=0.6$, and $\lambda_2=\lambda_3=0.4$. (d) The symmetry-breaking perturbation $\mathcal{H}^b_1$ leads to WSM, with $\kappa=0.5$.}
\label{fig4}
\end{figure}

\emph{Model B.}
We may construct a similar model which has the symmetry-protected DPs described by Eq.~(\ref{AFM_2}).
The lattice has a tetragonal structure of SG~102 ($P4'_2n'm$). As illustrated in Fig.~\ref{fig4}(a), there are four sublattices in a unit cell indexed by $(\tau_z,\sigma_z)=(\pm1,\pm1)$ with the basis vectors $\mathbf{t}_0=(1-\tau_z)(\frac{1}{4}\frac{1}{4}0)+(1-\sigma_z)(00\frac{1}{4})$. Similarly, each lattice site contains an $s$ orbital. The AFM interaction is strong with the ordering being along $\pm\hat{z}$ direction. $\mathcal{S}=e^{i\mathbf{k}\cdot\mathbf{d}}i\sigma_y\mathcal{K}$ with $\mathbf{d}=(\frac{1}{2}\frac{1}{2}0)$. The system is characterized by the symmetry generators $\{C_{4\hat{z}}|00\frac{1}{2}\}=\tau_xe^{i\pi\sigma_z/4}$ and $\{M_{\hat{x}}|\frac{1}{2}\frac{1}{2}\frac{1}{2}\}=\tau_x\sigma_y$. The generic and simplified model for the upper subsystem is
\begin{eqnarray}\label{AFM_model2}
\mathcal{H}_b &=& t\tau_x\cos\frac{k_z}{2}+t_1\tau_z\sin k_x\sin k_y
%\nonumber
\\
&&+\lambda_1\tau_z\left(\sigma_x\cos\frac{k_x}{2}\sin\frac{k_y}{2}+\sigma_y\sin\frac{k_x}{2}\cos\frac{k_y}{2}\right)
\nonumber
\\
&&+\lambda_2\tau_x\left(\sigma_x\sin\frac{k_x}{2}\cos\frac{k_y}{2}+\sigma_y\cos\frac{k_x}{2}\sin\frac{k_y}{2}\right)\cos\frac{k_z}{2}
\nonumber
\\
&&+\lambda_3\tau_y\left(\sigma_x\cos\frac{k_x}{2}\sin\frac{k_y}{2}+\sigma_y\sin\frac{k_x}{2}\cos\frac{k_y}{2}\right)\sin\frac{k_z}{2}
\nonumber
\end{eqnarray}
Here $t_i$ describes the hopping, $\lambda_i$ is SOC. Fig.~\ref{fig4}(c) shows energy band associated with $\mathcal{H}_b$, which features two inequivalent DPs at $Z$ and $A$ with linear dispersion. These two DPs are protected by $\mathcal{S}$, $\{C_{4\hat{z}}|00\frac{1}{2}\}$ and $\{M_{\hat{x}}|\frac{1}{2}\frac{1}{2}\frac{1}{2}\}$, which can be further seen by examining the
$\mathbf{k}\cdot\mathbf{p}$ model near these points. The representations of symmetry operations at $Z$ are, $\{C_{4\hat{z}}|00\frac{1}{2}\}=\tau_xe^{i\pi\sigma_z/4}$, $\{M_{\hat{x}}|\frac{1}{2}\frac{1}{2}\frac{1}{2}\}=\tau_x\sigma_y$ and $\mathcal{S}=i\tau_z\sigma_y\mathcal{K}$. Therefore the generic $\mathbf{k}\cdot\mathbf{p}$ Hamiltonian at $Z$ is
\begin{eqnarray}
\mathcal{H}_b^Z &=& (\zeta_0\sigma_x+\zeta_1\tau_z\sigma_y+\zeta_2\tau_y\sigma_y)k_x
\nonumber
\\
&&+(\zeta_0\sigma_y+\zeta_1\tau_z\sigma_x+\zeta_2\tau_y\sigma_x)k_y+\eta_1\tau_xk_z,
\end{eqnarray}
which leads to the dispersion
\begin{eqnarray}
E_{b,Z\pm}^2(\mathbf{k}) &=& (|\boldsymbol{\zeta}|^2+\zeta_0^2)(k_x^2+k_y^2)+|\boldsymbol{\eta}|^2k_z^2
\nonumber
\\
&&\pm2\sqrt{\zeta_0^2|\boldsymbol{\eta}|^2(k_x^2+k_y^2)k_z^2+4k_x^2k_y^2\zeta_0^2|\boldsymbol{\zeta}|^2}.
\end{eqnarray}
where $\boldsymbol{\zeta}\equiv(\zeta_1,\zeta_2,0)$ and $\boldsymbol{\eta}\equiv(\eta_1,0,0)$. When $|\boldsymbol{\zeta}|=|\zeta_0|$, one of the branches vanish on the line $k_x=k_y$, $k_z=0$. From model~(\ref{AFM_model2}), we have $\boldsymbol{\zeta}=(\frac{\lambda_1}{2},\frac{\lambda_3}{2},0)$ and $\zeta_0=0$, so a symmetry-protected DP exists at $Z$. A similar analysis applies to $A$.
Moreover, the symmetry-breaking topological phases of \emph{Model B} are studied in Table~\ref{table2}, where the perturbations are classified by their symmetry under the $C_{4v}$ point group. Take $\tau_x\sigma_z$ for example, it corresponds to a lattice distortion which breaks $\{M_{\hat{x}}|\frac{1}{2}\frac{1}{2}\frac{1}{2}\}$ but preserves $\{C_{4\hat{z}}|00\frac{1}{2}\}$. The reduced symmetry allows a term $\mathcal{H}^b_1=\kappa_1\sin(k_z/2)\tau_x\sigma_z$, which leads to a WSM as shown in Fig.~\ref{fig4}(d). It is worth mentioning that the lattice considered here is not a magnetic layer group, therefore by reducing the dimensionality we cannot access the 2D essential DSM phase.

\begin{table}[t]
\caption{Perturbations to the DP of \emph{Model B} in SG~102, classified by their symmetry under $C_{4v}$ point group~\cite{bradley1972}. The resulting insulating and semimetallic phases are indicated.}
\begin{center}\label{table2}
\renewcommand{\arraystretch}{1.12}
\begin{tabular*}{3.2in}
{@{\extracolsep{\fill}}ccc}
\hline
\hline
Reps & Perturbations & Phases
\\
\hline
$A_1$ & $1$ & DSM
\\
$A_2$ & $\tau_x\sigma_z$ & WSM
\\
$B_2$ & $\tau_z$ & NI
\\
      & $\tau_y$ & AFM TCI
\\
$E$ & $(\tau_x\sigma_x,\tau_x\sigma_y)$ & WSM
\\
\hline
\hline
\end{tabular*}
\end{center}
\end{table}

\emph{Discussion.}
The two models studied above provides an explicit extension of the relation between filling and essential nodal points~\cite{watanabe2015,po2016,watanabe2016,wieder2016b} to magnetic SGs. The complete study of such relation in 1651 magnetic SGs is left to future work. Note that in both models, there are two symmetry-inequivalent DPs which is symmetry tuned to the boundary between distinct TCI phases. It is possible to have an intrinsic magnetic DSM with only symmetry-equivalent DPs and no additional states at the Fermi energy, which separates NI and AFM TI with a nontrivial $Z_2$ index $\nu_0=1$~\cite{mong2010}. However, practically, constructing such a model requires a more complicated AFM ordering. Moreover, instead of $\mathcal{S}$, there also exists magnetic DSMs protected by $\bar{\Theta}\equiv C_n\Theta$~\cite{zhang2015}, which is beyond the scope of this paper.

Finally, we discuss some interesting physics and experimental consequence of the essential magnetic DSM. The nontrivial SSs~\cite{supplementary} and spin-orbit texture of Dirac cones can be directly measured by angle-resolved photoemission spectroscopy. The broken $\Theta$ and finite orbital magnetic moments of Fermi surface in magnetic DSMs may lead to a novel magnetopiezoelectric effect~\cite{varjas2016}. Macroscopically, $\mathcal{S}$-symmetry implies a topological magnetoelectric response $\partial P/\partial B=(\theta/2\pi)(e^2/h)$ with $\theta\neq0$ but not necessarily quantized on certain ferromagnetic surfaces with broken $\mathcal{S}$, provided the surface spectrum is gapped. Moreover, the magnetic fluctuations in this system may lead a dynamical axion field~\cite{li2010,wang2016a}. In terms of realistic materials, the actual existence of such phases in known materials remains an open question. However, unlike the band inversion induced topological DSM, here only the specific magnetic SG symmetry is needed for the essential magnetic DSM, which seems to be compatible with the narrow band width from AFM $d$- or $f$-orbitals. We are thus conservatively optimistic about the experimental prospects of 3D essential magnetic DSMs.

\begin{acknowledgments}
We acknowledge Naoto Nagaosa and Zhong Wang for valuable discussions. This work is supported by the National Thousand-Young-Talents Program; the National Key Research Program of China under Grant No.~2016YFA0300703; the Open Research Fund Program of the State Key Laboratory of Low-Dimensional Quantum Physics, through Contract No.~KF201606; and by Fudan University Initiative Scientific Research Program.
\end{acknowledgments}

\end{document}